\documentclass{article}
\usepackage{epsfig}
\usepackage[margin=1in]{geometry}

\newcommand{\beq}{\begin{equation}}
\newcommand{\eeq}{\end{equation}}
\newcommand{\beqa}{\begin{eqnarray}}
\newcommand{\eeqa}{\end{eqnarray}}
\newcommand{\lslash}[1]{#1\llap/}

\newcommand{\Eq}[1]{Eq.\ (\ref{#1})}
\newcommand{\Eqs}[2]{Eqs.\ (\ref{#1}) and (\ref{#2})}

\newcommand{\xRef}[1]{Ref.\ \cite{#1}}

\newcommand{\Fig}[1]{Fig.\ \ref{#1}}
\newcommand{\Section}[1]{Section\ \ref{#1}}

\title{
  Neutrino effective potential in a fermion and scalar background
}
  
\author{Jos\'e F. Nieves\footnote{nieves@ltp.uprrp.edu}\\
  Laboratory of Theoretical Physics, Department of Physics\\
  University of Puerto Rico, R\'{\i}o Piedras, Puerto Rico 00936
  \and\\[12pt]
  Sarira Sahu\footnote{sarira@nucleares.unam.mx}\\
  Instituto de Ciencias Nucleares\\
  Universidad Nacional Aut\'onoma de Mexico\\
  Circuito Exterior, C. U.\\
  A. Postal 70-543, 04510 Mexico DF, Mexico\\
}
\date{}

\begin{document}
\maketitle

\begin{abstract}
  We consider the neutrino self-energy
  in a background composed of a scalar particle and a fermion
  using a simple model for the coupling of the form $\lambda\bar f_R\nu_L\phi$.
  The results are useful in the context of Dark Matter-neutrino interaction
  models in which the scalar and/or fermion constitute the dark-matter.
  The corresponding formulas for models in which the scalar particle
  couples to two neutrinos via a coupling of the form
  $\lambda^{(\nu\nu\phi)}\bar\nu^c_R\nu_L\phi$
  are then obtained as a special case. In the presence of these interactions
  there can be new contributions to the neutrino
  effective potential and index of refraction
  in the context of neutrino collective oscillations in a supernova and
  in the Early Universe hot plasma before neutrino decoupling.
  The formulas obtained here can be used to estimate
  those effects and/or put limits on the model
  parameters based on the contribution that such interactions can have
  in those contexts. A particular feature of the results is that the
  contribution to the neutrino
  self-energy or effective potential in a neutrino background
  due to the $\nu\nu\phi$ coupling is proportional to the
  antineutrino-neutrino asymmetry $(n_{\bar\nu} - n_{\nu})$,
  in contrast to the standard $Z$ contribution which is
  proportional to $(n_{\nu} - n_{\bar\nu})$. Therefore the two contributions
  tend to cancel.  If the cancellation is significant, it is conceivable
  that the $O(1/m^4_\phi)$ terms give the dominant contribution.
  Alternatively, a limit can be set by requiring that the
  contribution of the $\nu\nu\phi$ interaction to the neutrino
  effective potential does not cancel the standard contribution in an
  appreciable way.
\end{abstract}

\section{Introduction}

The possible existence of complex scalars that interact with
neutrinos via a coupling of the form
\beq
\label{Lnunuphi}
L_{int} = \frac{1}{2}
\lambda^{(\nu\nu\phi)}\bar\nu^c_{R}\nu_{L}\phi + H.c\,,
\eeq
has been explored recently\cite{nunuphi,Farzan:2018gtr,%
Stephenson:1993rx,hm:2017mio,Heurtier:2016otg,%
Sawyer:2006ju,Pasquini:2015fjv,shao-feng,Brdar:2017kb}. Such scalars may
be produced in terrestrial neutrino experiments, and constraints
on their properties and interactions have been obtained from
particle physics, astrophysics, and cosmology
considerations in the works cited.

Here we note that such couplings would produce additional contributions
beyond the standard ones to the neutrino index of refraction and effective
potential when the neutrino propagates in a neutrino background. This occurs
in the environment of a supernova, where it is now well known that
the effect leads to the collective neutrino oscillations
and related phenomena\cite{Duan:2010bg},
and it can occur also in the hot plasma of the Early
Universe before the neutrinos decouple\cite{wong}. Therefore, further
constraints on the $\nu\nu\phi$ couplings and/or the $\phi$ properties
can be obtained by considering their effects in those contexts.

In a different line of development, models in which neutrinos
interact with a scalar and a fermion via a coupling of the form
\beq
\label{Lfnuphi}
L_{int} = \lambda\bar f_R\nu_L\phi + H.c.\,,
\eeq
have been considered recently in the context of Dark Matter-neutrino
interactions\cite{nu-dm1,nu-dm2}. Again, such interactions
produce additional contributions to the neutrino effective potential
when the neutrino propagates in a background of $\phi$ and $f$ particles.

In this paper we determine the effective potential
of a neutrino or antineutrino that propagates in such backgrounds.
To be specific, we calculate the neutrino self-energy in a background
of scalars $\phi$ and fermions $f$ due to the interaction 
given in \Eq{Lfnuphi}. From the self-energy,
the neutrino and antineutrino effective potential is then obtained.
The corresponding formulas for the case of the neutrino and scalar background,
with the couplings given in \Eq{Lnunuphi}, are obtained as the special
case in which $f_R \rightarrow \nu^c_R$. Our motivation is to provide a
uniform treatment and present the results in sufficiently general
way such that they can be applied to the situations described and also
adapted to others not considered here.
For example, models in which sterile neutrinos have \emph{secret} gauge
interactions of the form $\bar\nu_s\gamma^\mu \nu_s A^\prime_\mu$ have been
considered by several authors\cite{nusterilesecret}.
The formulas we obtain for the neutrino self-energy and effective potential,
including the case of an anisotropic background that we consider,
can be applied in the context of such models when a sterile neutrino
propagates in a background of sterile neutrinos and $A^\prime$ bosons,
with minor modifications.
  
In principle the scalar-neutrino couplings
can be of the form $\lambda^{(\nu\nu\phi)}_{ij}\bar\nu^c_{Ri}\nu_{Lj}$,
involving various neutrino species.
In writing \Eq{Lnunuphi} we are taking into account only the diagonal
neutrino coupling and assuming the presence of only one scalar field.
In the general case with more neutrino
species in the background and non-diagonal neutrino-$\phi$ couplings,
the density matrix formalism\cite{Duan:2010bg,wong} will come into play,
which is outside the scope of the present work. 
It is worth mentioning that, despite the simplification we are
making by considering only one neutrino type, the results
reveal some interesting and potentially important features that can
serve as a guide for considering the more general and realistic cases.
A particular noteworthy feature is that the contribution to the neutrino
self-energy or effective potential in the neutrino background
due to this coupling is proportional to
antineutrino-neutrino asymmetry $(n_{\bar\nu} - n_{\nu})$,
in contrast to the standard $Z$ contribution which is
proportional to $(n_{\nu} - n_{\bar\nu})$. Therefore the two contributions
tend to cancel. If the cancellation is significant, it is conceivable
that the $O(1/m^4_\phi)$ terms give the dominant contribution.
As an example application, we use these results to indicate
the constraint on $\lambda^{(\nu\nu\phi)}/m_\phi$ that
can be placed based on the contribution that such interactions can have
in the context of neutrino collective oscillations in a supernova.

In \Section{sec:dispersionrelations}, we review in a generic
way the conventions that we use to determine the neutrino
effective potential in a medium from the calculation of the self-energy.
The calculation of the self-energy of the neutrino propagating in the
background of fermions and scalars is undertaken in
\Section{sec:selfenergy}. We consider various cases
separately, depending on whether the background is isotropic or not,
and various limits of the momentum distribution functions of the
background particles.
In \Section{sec:nunuphi}, we use these results to calculate the effective
potential under different physical conditions for a $\nu\phi$ background.
We briefly summarize our results in \Section{sec:conclusions}.
  
\section{Dispersion relation and effective potential}
\label{sec:dispersionrelations}

\begin{figure}
\begin{center}
 \epsfig{file=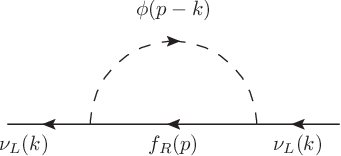,bbllx=222,bblly=351,bburx=386,bbury=426}
\end{center}
\caption[]{
  Neutrino self-energy diagram in a background of fermions$f$
  and scalars $\phi$.
  \label{fig:fnuphi}
}
\end{figure}

We consider the situation in which a neutrino propagates and interacts
with the $f$ and $\phi$ background particles as illustrated in
\Fig{fig:fnuphi}. For the purpose of determining the background
contribution to the neutrino self-energy and dispersion relation we
treat the neutrino as a massless chiral particle.
Denoting the neutrino momentum by $k^\mu = (\omega,\vec\kappa)$, and
the background contribution to the neutrino
self-energy by $\Sigma^{(T)}$,
the dispersion relations are obtained by solving
\beq
(\lslash{k} - \Sigma^{(T)})\psi_L = 0\,.
\eeq
The chirality of the neutrino interactions imply that
$\Sigma^{(T)}$ is of the form
\beq
\Sigma^{(T)} = V^\mu(\omega,\vec\kappa)\gamma_\mu\,.
\eeq
The dispersion relations are then given by the solutions of
\beq
\label{dispreleqV}
\omega - V^0 = \pm|\vec\kappa - \vec V|\,,
\eeq
which in general is an implicit equation for $\omega(\vec\kappa)$.
We consider two cases, according to whether the distribution functions
are isotropic or not.

\subsection{Isotropic case}

In this case $V^\mu$ is of the form
\beq
\label{Vab}
V_\mu = a(\omega,\kappa)k_\mu + b(\omega,\kappa)u_\mu\,,
\eeq
where $u^\mu$ is the velocity four-vector of the background.
In the frame in which the background is at rest, $u^\mu = (1,\vec 0)$,
which we adopt from now on in this case. In terms of $a$ and $b$,
\Eq{dispreleqV} becomes
\beq
\label{dispreleqab}
(1 - a)\omega - b = \pm(1 - a)\kappa\,.
\eeq
As already mentioned, in the most general case this is an implicit equation
for $\omega(\vec\kappa)$. On the other hand, for small $a$ and $b$,
the perturbative solutions around the vacuum solutions
$\omega_{\pm}(\vec\kappa) = \pm\kappa$ are to the lowest order
\beq
\omega_{\pm}(\vec\kappa) = \pm\kappa + b(\pm\kappa,\kappa)\,.
\eeq
The neutrino and antineutrino dispersion relations,
\beqa
\label{nubarnudisprel}
\omega_\nu(\vec\kappa) & = & \omega_{+}(\vec\kappa)\,,\nonumber\\
\omega_{\bar\nu}(\vec\kappa) & = & -\omega_{-}(-\vec\kappa)\,,
\eeqa
are then
\beq
\label{disprelVeff}
\omega^{(\nu,\bar\nu)}(\vec\kappa) = \kappa +
V^{(\nu,\bar\nu)}_{eff}(\vec\kappa)\,,
\eeq
where we have introduced the effective potentials $V^{(\nu,\bar\nu)}_{eff}$,
which in the present case are given explicitly by
\beqa
\label{Veffb}
V^{(\nu)}_{eff}(\vec\kappa) & = & b(\kappa,\kappa)\,,\nonumber\\
V^{(\bar\nu)}_{eff}(\vec\kappa) & = & -b(-\kappa,\kappa)\,.
\eeqa

\subsection{Anisotropic case}

In this case the $V^{\mu}$ cannot be decomposed in the
form given in \Eq{Vab}. We use \Eq{dispreleqV}
and assume that the equation can be solved perturbatively
around the vacuum solutions $\omega_{\pm}(\vec\kappa) = \pm\kappa$.
The solutions obtained in this way are then
\beqa
\omega_{+}(\vec\kappa) & = & \kappa + V^0(\kappa,\vec\kappa)
- \hat\kappa\cdot\vec V(\kappa,\vec\kappa)\,,\nonumber\\
\omega_{-}(\vec\kappa) & = & -\kappa + V^0(-\kappa,\vec\kappa)
+ \hat\kappa\cdot\vec V(-\kappa,\vec\kappa)\,.
\eeqa
Identifying the neutrino and antineutrino dispersion relations
by \Eq{nubarnudisprel}, they are given as in \Eq{disprelVeff},
where the effective potential in this case are
\beqa
\label{VeffV}
V^{(\nu)}_{eff}(\vec\kappa) & = & V^0(\kappa,\vec\kappa)
- \hat\kappa\cdot\vec V(\kappa,\vec\kappa)\,,\nonumber\\
V^{(\bar\nu)}_{eff}(\vec\kappa) & = & - V^0(-\kappa,-\vec\kappa)
+ \hat\kappa\cdot\vec V(-\kappa,-\vec\kappa)\,.
\eeqa
Of course in the isotropic case, in which $V^\mu$ can be decomposed
as in \Eq{Vab}, \Eq{VeffV} reduces to \Eq{Veffb}.

\section{Self-energy in the $f\phi$ background}
\label{sec:selfenergy}

In this section we consider a neutrino propagating
in a background of the fermion $f$ and complex scalar $\phi$
due to the interaction term in \Eq{Lfnuphi}. The starting point
is the calculation of the self-energy
via the diagram shown in \Fig{fig:fnuphi},
from which the dispersion relations are determined as indicated in
\Section{sec:dispersionrelations}.


The contribution of the diagram in \Fig{fig:fnuphi}
to the neutrino self-energy is given by
\beq
-iR\Sigma L = |\lambda|^2\int\,\frac{d^4p}{(2\pi)^4}
i\Delta^{(\phi)}_F(p - k) RiS^{(f)}_F(p)L \,,
\eeq
where
\beqa
iS^{(f)}_F(p) & = & iS^{(f)}_{F0}(p) -
\Gamma_{f}(p)\,,\nonumber\\
i\Delta^{(\phi)}_F(p) & = & i\Delta^{(\phi)}_{F0}(p) +
\Gamma_{\phi}(p)\,,\nonumber\\
\eeqa
with
\beqa
\Gamma_{f}(p) & = & \lslash{p}2\pi\delta(p^2 - m^2_f)
\eta_f(p)\,,
\nonumber\\
\Gamma_{\phi}(p) & = & 2\pi\delta(p^2 - m^2_\phi)\eta_\phi(p)\,.
\eeqa
Using the label $x$ to stand for either $f$ or $\phi$,
the functions $\eta_x(p)$ are given by
\beq
\eta_x(p) = \theta(p^0)f_x(p^0,\vec p) +
\theta(-p^0)f_{\bar x}(-p^0,-\vec p)\,,
\eeq
where  $f_{x,\bar x}(E_p,\vec p)$ are the momentum distribution functions
of the background particles and antiparticles,
which may not necessarily be isotropic.
For an isotropic thermal background, the distribution functions are
\beqa
\label{thermaldist}
f_{f,\bar f}(E_f) & = &
\frac{1}{e^{\beta E_f \mp \alpha_f} + 1}\,,\nonumber\\
f_{\phi,\bar\phi}(E_\phi) & = & 
\frac{1}{e^{\beta E_\phi \mp \alpha_\phi} -1}\,.
\eeqa

Discarding the pure vacuum contribution to $\Sigma$ and denoting
by $\Sigma^{(f,\phi)}$ the background-dependent part, we write
\beq
\Sigma^{(f,\phi)} = \Sigma^{(f)} + \Sigma^{(\phi)}\,,
\eeq
where
\beqa
\label{Sigmaf}
\Sigma^{(f)} & = & -|\lambda|^2\int\frac{d^4p}{(2\pi)^3}
\frac{\lslash{p}}{(p - k)^2 - m^2_\phi}
\delta(p^2 - m^2_f)\eta_f(p)\,,\\
\label{Sigmaphi}
\Sigma^{(\phi)} & = & |\lambda|^2\int\frac{d^4p}{(2\pi)^3}
\frac{(\lslash{p} + \lslash{k})}{(p + k)^2 - m^2_f}
\delta(p^2 - m^2_\phi)\eta_\phi(p)\,.
\eeqa
We will evaluate $\Sigma^{(f,\phi)}$ in the following limiting cases.
In the first case, to which we refer as the \emph{high temperature}
limit, we assume that the $f$ and $\phi$ backgrounds are isotropic,
and also extremely relativistic so that their respective
masses $m_{f,\phi}$ and chemical potentials $\alpha_{f,\phi}$
can be taken to be zero in \Eqs{Sigmaf}{Sigmaphi}.
In the second case we also assume the isotropic
thermal distribution functions, but with
the temperature and chemical potentials being such that the integrand
in \Eqs{Sigmaf}{Sigmaphi} can be approximated by expanding them
in powers of
\beq
\label{Delta}
\Delta \equiv m^2_\phi - m^2_f\,.
\eeq
We refer to this as the \emph{heavy background} limit.
This also includes the case in which either $f$ or $\phi$ is massless,
provided that the other particle is heavy enough.
Finally we will consider a variant of the heavy background case,
in which the distribution functions are not isotropic.
By taking $f_R = \nu^c_R$ we can apply the results of any case,
depending on whether $\phi$ is a light or heavy particle
in the sense specified above, to the model in which $\phi$ couples to
$\bar\nu^c_R\nu_L$ and consider situations in which a neutrino
propagates in a neutrino background.
Following the discussion in \Section{sec:dispersionrelations},
each $\Sigma^{(x)}$ is of the form
\beq
\label{SigmaxV}
\Sigma^{(x)} = V^{(x)\mu}\gamma_\mu\,,
\eeq
and
\beq
V_\mu = V^{(f)}_\mu + V^{(\phi)}_\mu\,.
\eeq
When the background is isotropic each $V^{(x)\mu}$ can be decomposed as
\beq
\label{Sigmaxab}
V^{(x)}_\mu = a^{(x)}k_\mu + b^{(x)}u_\mu\,.
\eeq

\subsection{High temperature limit}
\label{subsec:fphihighT}

Here we assume that $T$ and $\kappa$ are sufficiently large
such that $f$ and $\phi$ can be taken to be massless.
As we have already stated above, we then approximate the
integrals in \Eqs{Sigmaf}{Sigmaphi} by setting $m_{f,\phi}$ and
chemical potentials $\alpha_{f,\phi}$ to zero. In this case
\beqa
\label{SigmafHT}
\Sigma^{(f)} & = & |\lambda|^2\int\frac{d^4p}{(2\pi)^3}\delta(p^2)
\frac{\lslash{p}}{(p + k)^2}
\frac{1}{e^{|\beta p\cdot u|} + 1}\,,\\
\label{SigmaphiHT}
\Sigma^{(\phi)} & = & |\lambda|^2\int\frac{d^4p}{(2\pi)^3}\delta(p^2)
\frac{(\lslash{p} + \lslash{k})}{(p + k)^2}
\frac{1}{e^{|\beta p\cdot u|} - 1}\,.
\eeqa
We will restrict ourselves to the dominant contribution in powers of $T$.
In this case the term with the factor of $\lslash{k}$ in the integrand
of \Eq{SigmaphiHT} can be discarded and we can write
\beq
\Sigma^{(x)} = |\lambda|^2 I^{(x)}_\mu\gamma^\mu\,,
\eeq
where
\beq
\label{integralsIER}
I^{(f,\phi)}_\mu = \int\frac{d^4p}{(2\pi)^3}\delta(p^2)
\frac{p_\mu}{(p + k)^2}\frac{1}{e^{|\beta p\cdot u|} \pm 1}\,.
\eeq
The integrals in \Eq{integralsIER} are the same ones
evaluated by Weldon for chiral fermions in a
non-Abelian gauge theory\cite{weldon-fermions} and also used
in a previous discussion of the $\nu\nu\phi$ interaction
model in \xRef{nu-inmedium}. Borrowing from those calculations,
the results for $\Sigma^{(f,\phi)}$ can be written
in the form of \Eq{Sigmaxab} with
\beqa
\label{abfphier}
a^{(\phi)} = 2a^{(f)} & = & -\frac{|\lambda|^2 T^2}{24\kappa^2}
\left[1 - \frac{\omega}{2\kappa}I\right]\nonumber\\
b^{(\phi)} = 2b^{(f)} & = & \frac{|\lambda|^2 T^2}{24\kappa}
\left[\frac{\omega}{\kappa} - 
\frac{1}{2}\left(\frac{\omega^2}{\kappa^2} -1 \right)I\right]\,,
\eeqa
%
%
%
where
\beq
I \equiv \ln\left|\frac{\omega + \kappa}{\omega - \kappa}\right|\,.
\eeq

The total background-dependent part of the self-energy is then
\beq
\label{Sigmafphiab}
\Sigma^{(f\phi)} = a^{(f\phi)}\lslash{k} + b^{(f\phi)}\lslash{u}\,,
\eeq
with
\beqa
\label{abf+phier}
a^{(f\phi)} & = & -\frac{M^2}{\kappa^2}
\left[1 - \frac{\omega}{2\kappa}I\right]\nonumber\\
b^{(f\phi)} & = & \frac{M^2}{\kappa}
\left[\frac{\omega}{\kappa} - 
\frac{1}{2}\left(\frac{\omega^2}{\kappa^2} -1 \right)I\right]\,,
\eeqa
where
\beq
M^2 = \frac{|\lambda|^2 T^2}{16}\,.
\eeq
The corresponding dispersion relations in this case
are given by the solutions of \Eq{dispreleqab},
with the coefficients $a,b$ given by \Eq{abf+phier}.
They are similar to the dispersion relations obtained by
Weldon\cite{weldon-fermions} for a chiral fermion in a
non-Abelian gauge theory, which in particular have
$\omega(\kappa \rightarrow 0) = \pm M$, and other interesting
features\cite{weldon:particlesandholes,klimov}.

\subsection{Heavy background}

In this case we assume that either $f$ of $\phi$ is a heavy particle,
in the sense that its mass is sufficiently larger than $T$, $\omega$
and $\kappa$, such that the denominators in the integrand
in \Eqs{Sigmaf}{Sigmaphi} can be expanded in the form
\beqa
\frac{1}{(p + k)^2 - m^2_f} & = & \frac{1}{\Delta} -
\frac{2p\cdot k + k^2}{\Delta^2}\,,\nonumber\\
\frac{1}{(p - k)^2 - m^2_\phi} & = & -\frac{1}{\Delta} +
\frac{2p\cdot k - k^2}{\Delta^2}\,,
\eeqa
where $\Delta$ is defined in \Eq{Delta}.
Thus, $\Sigma^{(f,\phi)}$ are given as in \Eq{SigmaxV} with
%
%
%
\beqa
V^{(f)}_\mu(\omega,\vec\kappa) & = & |\lambda|^2
\left(\frac{1}{\Delta} + \frac{k^2}{\Delta^2}\right)L^{(f)}_\mu
-\frac{2|\lambda|^2}{\Delta^2} L^{(f)}_{\mu\nu}k^\nu
\,,\nonumber\\
V^{(\phi)}_\mu(\omega,\vec\kappa) & = & |\lambda|^2
\left(\frac{1}{\Delta} - \frac{k^2}{\Delta^2}\right)L^{(\phi)}_\mu
- \frac{2|\lambda|^2}{\Delta^2}L^{(\phi)}_\nu k^\nu k_\mu\nonumber\\
&&\mbox{} + |\lambda|^2
\left(\frac{1}{\Delta} - \frac{k^2}{\Delta^2}\right)L^{(\phi)}k_\mu
- \frac{2|\lambda|^2}{\Delta^2} L^{(\phi)}_{\mu\nu}k^\nu\,,
\eeqa
where
\beq
\label{defLintegrals}
(L^{(x)},L^{(x)}_\mu,L^{(x)}_{\mu\nu}) =
\int\,\frac{d^4p}{(2\pi)^3}\delta(p^2 - m^2_x)
\eta_x(p)\left\{1,p_\mu,p_\mu p_\nu\right\}\,.
\eeq
In order to obtain the effective potential from \Eq{VeffV},
we use the fact that, for $\omega = \kappa$, we have
\beq
k^\mu = \kappa n^\mu\,,
\eeq
where
\beq
n^\mu = (1,\hat\kappa)\,,\qquad n^2 = 0\,.
\eeq
From \Eq{VeffV} the effective potentials are then given by
%
%
\beq
\label{VeffLn}
V^{(\nu,\bar\nu)}_{eff}(\vec\kappa) = \pm \frac{|\lambda|^2}{\Delta}
\left(L^{(f)}_1 + L^{(\phi)}_1\right) - \frac{2|\lambda|^2}{\Delta^2}
\left(L^{(f)}_2 + L^{(\phi)}_2\right)\,,
\eeq
where the upper(lower) sign corresponds to the neutrino(antineutrino),
respectively.
%
%
In \Eq{VeffLn} the coefficients $L^{(x)}_n$ are defined as
\beqa
L^{(x)}_1 & \equiv & L^{(x)}_\mu n^\mu \,,\nonumber\\
L^{(x)}_2 & \equiv & L^{(x)}_{\mu\nu} n^\mu n^\nu\,,
\eeqa
and using the definitions in \Eq{defLintegrals},
the following formula follows
\beq
L^{(x)}_n = \frac{1}{2}
\int\,\frac{d^3p}{(2\pi)^3} E^{n - 1}_x
\left(1 - \hat\kappa\cdot\vec v_{\vec p}\right)^n
\left(f_x + (-1)^n f_{\bar x}\right)\,.
\eeq

For isotropic momentum distributions functions we can
set $\vec p\rightarrow 0$ and $p^i p^j\rightarrow \frac{1}{3}{\vec p}^{\,2}$
in the integrands of $L^{(x)}_1$ and $L^{(x)}_2$, respectively, and the
formulas reduce in that case to
%
\beqa
\label{LJrelations}
L^{(x)}_1 & = & J^{(x)}_1\,,\nonumber\\
L^{(x)}_2 & = & \frac{4}{3}J^{(x)}_2 - \frac{1}{3}m^2_x J^{(x)}_0\,,
\eeqa
%
%
%
%
where
%
%
%
%
\beq
\label{Jn}
J^{(x)}_n = 
\frac{1}{2}
\int\,\frac{d^3p}{(2\pi)^3} E^{n - 1}_x\left(f_x + (-1)^n f_{\bar x}\right)\,.
\eeq
%
%
%
%
%
%
%
%
Using \Eq{LJrelations}, the formulas given in \Eq{VeffLn}
for the neutrino and antineutrino effective potential then reduce to
%
%
\beq
\label{VeffJn}
V^{(\nu,\bar\nu)}_{eff}(\vec\kappa) = \pm \frac{|\lambda|^2}{\Delta}
\left(J^{(f)}_1 + J^{(\phi)}_1\right) +
\frac{2|\lambda|^2\kappa}{3\Delta^2}\left(
m^2_f J^{(f)}_0 + m^2_\phi J^{(\phi)}_0 - 4 J^{(f)}_2 - 4 J^{(\phi)}_2
\right)\,,
\eeq
in the isotropic case. Similarly to \Eq{VeffLn},
in \Eq{VeffJn} the the upper(lower) sign corresponds
to the neutrino(antineutrino), respectively.
Introducing the number densities
\beq
\label{numberdensities}
n_{x,\bar x} = g_x\int\,\frac{d^3p}{(2\pi)^3}f_{x,\bar x}\,,
\eeq
%
%
%
%
%
%
we have
\beq
\label{J1}
J^{(x)}_1 = \frac{1}{2g_x}(n_x - n_{\bar x})\,,
\eeq
where
\beq
g_f = 2\,,\quad g_\phi = 1\,.
\eeq
In anticipation to the application of these formulas in the next section
we must remember that for chiral neutrinos
\beq
\label{gfactornu}
g_\nu = 1\,.
\eeq
It is straightforward to obtain explicit formulas for the integrals
$J^{(x)}_{0,2}$ in a number of cases. For example,
\beqa
\label{Jnexamples}
J^{(x)}_n = \frac{1}{2g_x}\left(n_x + (-1)^n n_{\bar x}\right)\left\{
\begin{array}{cl}
m^{n - 1}_x &
(\mbox{NR limit})\\[12pt]
\frac{1}{2}(n+1)!\beta^{1 - n} & \mbox{(Maxwell-Boltzman ER limit)}
\end{array}
\right.
\eeqa
in the non-relativistic (NR) or the Maxwell-Boltzman
extremely-relativistic (ER) limits.

\section{$\nu\phi$ background}
\label{sec:nunuphi}

\begin{figure}
\begin{center}
\epsfig{file=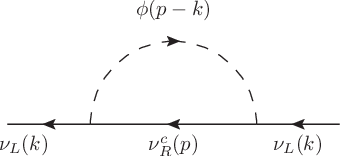,bbllx=222,bblly=351,bburx=386,bbury=426}
\end{center}
\caption[] {Neutrino self-energy diagram in a background of neutrinos
  and scalars $\phi$.}
\label{fig:nunuphi}
\end{figure}

In this section we use the results obtained in \Section{sec:selfenergy}
to discuss the effects of the interaction given in \Eq{Lnunuphi} 
on a neutrino propagating in a neutrino-$\phi$ background.
It is useful to remember that in the relevant diagram for this case, as shown
in \Fig{fig:nunuphi}, the internal fermion that corresponds to the
$f_R$ fermion line in \Fig{fig:fnuphi} is the antineutrino $\nu^c_R$.
As a result of this, the formulas obtained in \Section{sec:selfenergy}
can be adapted to this case provided we identify
$f,\bar f\rightarrow \bar\nu,\nu$ in the labels of the various
physical quantities that refer to the background particles,
such as the chemical potentials, particle number densities.
In addition, we have to remember that for chiral neutrinos
we must use $g_\nu = 1$ in the formula for $J^{(\bar\nu)}_1$ given
in \Eq{J1}, as stated in \Eq{gfactornu}.

\subsection{High temperature limit}

Denoting the background-dependent part of the self-energy in this case
by $\Sigma^{(\bar\nu\phi)}$ and writing
\beq
\label{Sigmaparametrization}
\Sigma^{(\bar\nu\phi)} = a^{(\bar\nu\phi)}\lslash{k} +
b^{(\bar\nu\phi)}\lslash{u}\,,
\eeq
the coefficients are given by the same expressions of \Eq{abf+phier},
with
%
%
%
%
%
%
\beq
M^2 \equiv \frac{1}{16} |\lambda^{(\nu\nu\phi)}|^2 T^2 \,.
\eeq
These are the results used also in a previous discussion of
the $\nu\nu\phi$ interaction model in \xRef{nu-inmedium}.
As already mentioned in \Section{subsec:fphihighT},
the dispersion relation for the neutrino in this case
is similar to the dispersion relation obtained by
Weldon\cite{weldon-fermions} for a chiral fermion in a
non-Abelian gauge theory.
However, this case is not the one that is relevant to the
situations where $m_\phi$, although assumed to be
below the electroweak breaking scale, is assumed to be $\sim GeV$,
such as those considered in \xRef{nunuphi}.
Thus we turn the attention to the case in which $\phi$ is
sufficiently heavy such that its background density is negligible.

\subsection{Heavy $\phi$ limit}

For illustrative purposes we consider in some detail the isotropic case.
From \Eq{Veffb}, neglecting the $\phi$ background terms
and putting $m_\nu  = 0$, the $\nu\phi$ background contribution to the
effective potentials is then
\beq
\label{VeffJn-nubackground}
\left(V^{(\nu,\bar\nu)}_{eff}(\vec\kappa)\right)_{\nu\nu\phi} =
\pm \frac{|\lambda^{(\nu\nu\phi)}|^2}{m^2_\phi}J^{(\bar\nu)}_1 -
\frac{8|\lambda^{(\nu\nu\phi)}|^2\kappa}{3m^4_\phi}J^{(\bar\nu)}_2\,.
\eeq
The full background-dependent effective potential is then
\beq
V^{(\nu,\bar\nu)}_{eff}(\vec\kappa) = 
\left(V^{(\nu,\bar\nu)}_{eff}(\vec\kappa)\right)_{\nu\nu\phi} +
\left(V^{(\nu,\bar\nu)}_{eff}(\vec\kappa)\right)_{Z}\,,
\eeq
where $\left(V^{(\nu,\bar\nu)}_{eff}\right)_{Z}$
is the \emph{standard} neutrino-background
contribution from the $Z$ exchange and tadpole diagrams
\beq
\label{Zcontribution}
\left(V^{(\nu,\bar\nu)}_{eff}(\vec\kappa)\right)_{Z} =
\pm 4\sqrt{2} G_F J^{(\nu)}_1\,.
\eeq
\Eq{Zcontribution} follows, from example, 
from the formulas obtained in \xRef{dnt}.
As shown there, the contribution of those two diagrams to the
neutrino self-energy can be written in the form of
\Eqs{SigmaxV}{Sigmaxab}, with
\beqa
\left(b^{(Z)}\right)_{tadpole}\ & = & 4\sqrt{2}G_F\sum_f X_\nu J^{(f)}_1\,,
\nonumber\\
\left(b^{(Z)}\right)_{Z-exchange} & = & 2\sqrt{2}G_F J^{(\nu)}_1\,,
\eeqa
where the sum runs over all the fermion species in the background
and $X_f$ is the (vector) neutral-current coupling of the fermion.
In \Eq{Zcontribution} we are including only the neutrino background
contribution, and using $X_\nu = \frac{1}{2}$.
Remembering \Eqs{J1}{gfactornu},
\beq
J^{(\bar\nu)}_1 = -J^{(\nu)}_1 = \frac{1}{2}(n_\nu - n_{\bar\nu})\,,
\eeq
we then have
\beq
\label{Veffnuphi+Z}
V^{(\nu,\bar\nu)}_{eff}(\vec\kappa) = \pm 
\left(2\sqrt{2}G_F - \frac{|\lambda^{(\nu\nu\phi)}|^2}{2m^2_\phi}\right)
(n_\nu - n_{\bar\nu}) -
\frac{8|\lambda^{(\nu\nu\phi)}|^2\kappa}{3m^4_\phi}J^{(\bar\nu)}_2\,.
\eeq

\subsection{Discussion}
\label{sec:discussion}

In a core collapse supernova, neutrinos are trapped in the neutrino
sphere and slowly diffuse out. These neutrinos have self-interactions
that lead what is known as collective neutrino
oscillations\cite{Duan:2010bg}.
These phenomena can also occur in the hot plasma of the Early
Universe\cite{wong} before the neutrinos decouple. In the presence of the
type of neutrino-scalar interaction we have considered, the collective
oscillations could be modified.

A striking and surprising result is that the standard
contribution $\left(V^{(\nu,\bar\nu)}_{eff}\right)_{Z}$ on one hand,
and $\left(V^{(\nu,\bar\nu)}_{eff}\right)_{\nu\nu\phi}$ on the other,
have opposite sign. Namely, they are proportional to
$J^{(\nu)}_1 = \frac{1}{2}(n_\nu - n_{\bar\nu})$ and
$J^{(\bar\nu)}_1 = \frac{1}{2}(n_{\bar\nu} - n_{\nu})$, respectively.
Thus, they tend to cancel.
A limit can be set by requiring that the
contribution of the $\nu\nu\phi$ interaction to the neutrino
self-energy does not cancel the standard contribution in an
appreciable way. In symbols we can express this as
\beq
\left(V^{(\nu)}_{eff}\right)_{\nu\nu\phi} < \left(V^{(\nu)}_{eff}\right)_{Z}\,,
\eeq
which requires
\beq
m_\phi > \frac{\lambda^{(\nu\nu\phi)}}{g}m_Z\,.
\eeq
Therefore, if $\lambda^{(\nu\nu\phi)} \sim O(1)$, this would imply
$m_\phi$ must be larger than $\sim 3m_Z$. In order to have
$m_\phi \sim O(GeV)$, a small coupling
$\lambda^{(\nu\nu\phi)} \sim 10^{-2} - 10^{-3}$ would be required.

Another possibility is that the first term in \Eq{Veffnuphi+Z},
proportional to the neutrino-antineutrino asymmetry is small,
either because the factor in parenthesis and/or because the
asymmetry is small. In such a case, the $O(1/m^4_\phi)$ term
in effective potential
\beq
V^{(\nu,\bar\nu)}_{eff}(\vec\kappa) = -
\frac{8|\lambda^{(\nu\nu\phi)}|^2\kappa}{3m^4_\phi}J^{(\bar\nu)}_2\,.
\eeq
would give the dominant contribution. For example, taking
the Maxwell-Boltzman limit of the neutrino background
from \Eq{Jnexamples}, this would give
\beq
V^{(\nu,\bar\nu)}_{eff}(\vec\kappa) = -
\frac{2|\lambda^{(\nu\nu\phi)}|^2\kappa}{m^4_\phi}
\frac{(n_\nu + n_{\bar\nu})}{\beta}\,,
\eeq
which is the same for the neutrino and antineutrino,
and independent of the neutrino-antineutrino asymmetry in the background.

Although we have focused above on the case of isotropic
thermal distributions of the neutrino background, similar considerations
apply to the case of anisotropic distributions as well.

\section{Conclusions}
\label{sec:conclusions}

The existence of scalars that interact with neutrinos via
couplings of the form $\lambda^{(\nu\nu\phi)}\bar\nu^c_{R}\nu_{L}\phi$
would produce additional contributions to the neutrino effective potential
beyond the standard ones when the neutrino propagates in a
neutrino background. This can occur in the environment
of a supernova, and in the Early Universe hot plasma before
neutrino decoupling.

Motivated by studying the possible effects of such
interactions in those contexts, we have considered a simple model in
which neutrino interacts with a scalar and a fermion, and calculated
the effective potential for a neutrino propagating
in a background of those particles.  
The results are useful in the context of Dark Matter-neutrino interaction
models in which the scalar and/or fermion constitute the dark-matter,
and are also applicable to the situations mentioned in which the
fermion background is a neutrino background.
We obtained the expressions for the neutrino effective potential
that can be applied to different situations and background conditions.

As a specific application we considered the case of the neutrino background,
with the scalar particle being sufficiently heavy so that their background
density is negligible. A noteworthy result in this case is that
the effective potential experienced by the propagating neutrino
in this background has the opposite sign to the effective potential
due to the standard contribution from the Z-boson.
By assuming that the standard model contribution dominates over 
the $\nu\nu\phi$ contribution, for example in the neutrino sphere
of a core collapse supernova, we have obtained limit on the parameter
$\lambda^{(\nu\nu\phi)}/m_{\phi}$.

Although for definiteness we have restricted ourselves to consider
the diagonal scalar-neutrino coupling, the results already
reveal some potentially important features, for example the sign difference
mentioned above, that can serve as a guide for
considering more realistic or complicated models involving,
for example, more than one scalar particle or off-diagonal
neutrino couplings.


\begin{thebibliography}{99}

\bibitem{nunuphi}
  Jeffrey M. Berryman, Andr\'e de Gouvea, Kevin J. Kelly and Yue Zhang,
  \emph{
    Lepton-Number-Charged Scalars and Neutrino Beamstrahlung
  },
  arxiv:1802.00009 (jan 2018)
 
\bibitem{Farzan:2018gtr} 
  Y.~Farzan, M.~Lindner, W.~Rodejohann and X.~J.~Xu,
  \emph{Probing neutrino coupling to a light scalar with
    coherent neutrino scattering},
  JHEP {\bf 1805}, 066 (2018)
  [arXiv:1802.05171 [hep-ph]].

\bibitem{Stephenson:1993rx} 
  G.~J.~Stephenson, Jr. and J.~T.~Goldman,
  \emph{Observable consequences of a scalar boson coupled only to neutrinos},
  hep-ph/9309308.

\bibitem{hm:2017mio} 
  C.~Boehm, A.~Olivares-Del Campo, S.~Palomares-Ruiz and S.~Pascoli,
  \emph{Phenomenology of a Neutrino-DM Coupling: The Scalar Case},
  arXiv:1705.03692 [hep-ph].

\bibitem{Heurtier:2016otg} 
  L.~Heurtier and Y.~Zhang,
  \emph{Supernova Constraints on Massive (Pseudo)Scalar Coupling to Neutrinos},
  JCAP {\bf 1702}, no. 02, 042 (2017)
  [arXiv:1609.05882 [hep-ph]].

\bibitem{Sawyer:2006ju} 
  R.~F.~Sawyer,
  \emph{Bulk viscosity of a gas of neutrinos and coupled scalar particles,
    in the era of recombination},
  Phys.\ Rev.\ D {\bf 74}, 043527 (2006)
  [astro-ph/0601525].

\bibitem{Pasquini:2015fjv} 
  P.~S.~Pasquini and O.~L.~G.~Peres,
  \emph{Bounds on Neutrino-Scalar Yukawa Coupling},
  Phys.\ Rev.\ D {\bf 93}, no. 5, 053007 (2016);
  Erratum: [Phys.\ Rev.\ D {\bf 93}, no. 7, 079902 (2016)]
  [arXiv:1511.01811 [hep-ph]].

  \bibitem{shao-feng}
  Shao-Feng Ge, Manfred Lindner and Werner Rodejohann,
  \emph{Atmospheric Trident Production for Probing New Physics},
  Phys.\ Lett.\ B 772, 164 (2017);
  arxiv:1702.02617

\bibitem{Brdar:2017kb}
  Vedran Brdar, Joachim Kopp, Jia Liu, Pascal Prass and Xiao-Ping Wang,
  \emph{Fuzzy dark matter and nonstandard neutrino interactions},
  Phys.\ Rev.\ D \textbf{97}, 043001 (2018)
  [arxiv: 1705.09455].

\bibitem{Duan:2010bg} 
  See for example,
  H.~Duan, G.~M.~Fuller and Y.~Z.~Qian,
  \emph{Collective Neutrino Oscillations},
  Ann.\ Rev.\ Nucl.\ Part.\ Sci.\  {\bf 60}, 569 (2010)
  [arXiv:1001.2799 [hep-ph]],
  and references therein


\bibitem{wong}
  See for example,
  Yvonne Y. Y. Wong,
  \emph{
    Analytical treatment of neutrino asymmetry equilibration
    from flavor oscillations in the early universe
  }
  Phys.\ Rev.\ D 66, 025015  (2002), and references therein.

\bibitem{nu-dm1}
  Reinard Primulando and Patipan Uttayarat,
  \emph{
    Dark Matter-Neutrino Interaction in Light of Collider
    and Neutrino Telescope Data
  },
  arxiv:1710.08567 (june 2018)

\bibitem{nu-dm2}
  Tarso Franarin, Malcolm Fairbairn and Jonathan H. Davis,
  \emph{
    JUNO Sensitivity to Resonant Absorption of Galactic Supernova
    Neutrinos by Dark Matter
  },
  arxiv:1806.05015 (june 2018)

\bibitem{nusterilesecret}
  Xiaoyong Chu, Basudeb Dasgupta, Mona Dentler, Joachim Kopp, Ninetta Saviano,
  \emph{Sterile Neutrinos with Secret Interactions -- Cosmological Discord?},
  https://arxiv.org/abs/1806.10629 (June 2018) and references therein.

\bibitem{weldon-fermions}
  H. Arthur Weldon,
  Phys.\ Rev.\ D 26, 2789 (1982)

\bibitem{nu-inmedium}
  Jos\'e F. Nieves,
  \emph{Neutrinos in a  medium},
  Phys.\ Rev.\ D40, 866 (1989)
  
\bibitem{weldon:particlesandholes}
  H. Arthur Weldon, Physica\ A\ 158, 169 (1989)

\bibitem{klimov}
  V.V. Klimov, Soy.\ J.\ Nucl.\ Phys. 33, 934 (1981)

\bibitem{dnt}
  Juan Carlos D'Olivo, Jos\'e F. Nieves and Manuel Torres,
  Phys.\ Rev.\ D 46, 1172 (1992).

\end{thebibliography}
\end{document}